\documentclass[multphys,vecphys]{svmult}

\usepackage{makeidx}         
\usepackage{graphicx}        
\usepackage{multicol}        
\usepackage[bottom]{footmisc}

\makeindex             

\begin{document}

\title*{Direct detection of exo-planets: GQ Lupi\thanks{Proceedings ESO Workshop on
{\it The power of optical/IR interferometry}}}
\author{Ralph Neuh\"auser\inst{1}\and Eike Guenther\inst{2}\and Peter Hauschildt\inst{3}}
\institute{Astrophysikalisches Institut und Universit\"ats-Sternwarte, Schillerg\"a\ss chen 2, D-07745 Jena, Germany
\texttt{rne@astro.uni-jena.de}
\and Th\"uringer Landessternwarte Tautenburg, D-07778 Tautenburg, Germany
\and Hamburger Sternwarte, Gojenbergsweg 112, D-21029 Hamburg, Germany}
%
%
\maketitle

\section{Introduction: GQ Lupi and its companion}

Since several years, we have been searching for sub-stellar
companions around young (up to 100 Myrs) nearby (up to 150 pc)
stars, both brown dwarfs and giant planets. Young sub-stellar
objects are self-luminous due to ongoing contraction and
accretion, so that young stars are good targets.
We have found several brown dwarf companion candidates and
confirmed three of them by both proper motion and spectroscopy:
TWA-5 B, HR 7329 B, and GSC 8047 B, all being few Myr young
late M-type brown dwarfs.

With K-band imaging using VLT/NaCo, Subaru/CIAO, and HST/PC, we detected a
6 mag fainter object $0.7 ^{\prime \prime}$ west of the classical T Tauri
star GQ Lup, which is a clearly co-moving companion ($\ge 10~\sigma$), 
but orbital motion was not yet detectable. The NaCo K-band spectrum yielded
$\sim$ L1-2 (M9-L4) as spectral type. At $140 \pm 50$ pc distance (Lupus I cloud),
it can be placed into the H-R diagram.
For more details, see Neuh\"auser et al. (2005).
According to our own calculations following Wuchterl \& Tscharnuter
(2003), it has 1 to 3~M$_{\rm jup}$, but according to Baraffe et al. (2002) and
Burrows et al. (1997) models, it is anywhere between 3 and 42~M$_{\rm jup}$.
The latter models are not applicable for the young GQ Lup and its companion,
while the former does take into account the formation and collapse;
the Wuchterl \& Tscharnuter (2003) convection model is calibrated on the Sun.

\section{Comparison with model atmospheres}

We use the new so-called GAIA-dusty grid (Brott \& Hauschildt et al., in preparation),
which is an updated version of the models from Allard et al. (2001). 
The updates include improved molecular dissociation constants as well as more dust
species and their opacities. In addition, the models were computed for a 
convective mixing length parameter of 1.5 times the pressure scale height H$_{\rm p}$.
It uses spherical symmetry, which is most important for 
young and sub-stellar objects as GQ Lup A and b with low gravities.
The Allard et al. (2001) AMES-dusty sometimes had problems at very
low gravity.

We compare our K-band spectrum (see Neuh\"auser et al. 2005)
with the GAIA grid for temperatures T$_{\rm eff} = 2000$ 
and 2900 K and for gravities of $\log~g=0, 2$, and 4 (g in cgs units).
See figure 1 for the comparison.

\begin{figure}
\centering
\includegraphics[width=\columnwidth]{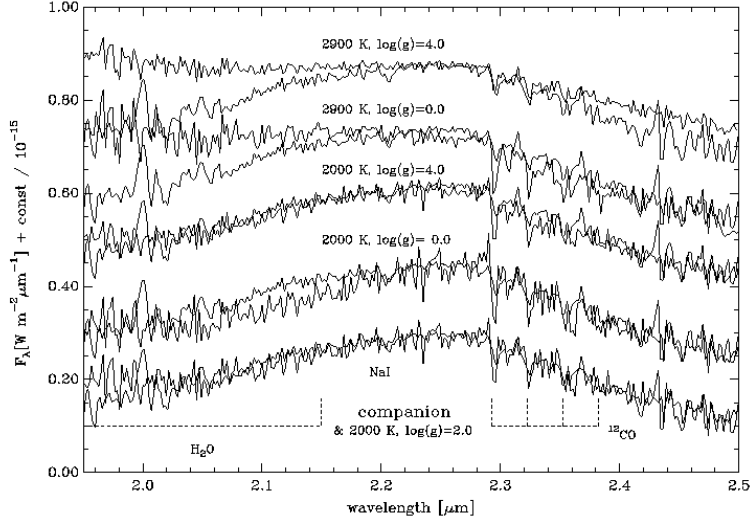}
%
%
\caption{Comparison of our GAIA-dusty model atmosphere grid with
the observed spectrum overplotted. The high T=2900 K in the upper two spectra
do not reproduce the water steam absorption in the blue.
The model with T=2000 K and log g =2 fits best (bottom), 
see text. H$_{2}$O, Na, and 12CO are indicated.}
\end{figure}

The model spectrum with T$_{\rm eff} = 2000$ K is much better than 
the hotter temperature (where there is no water vapour absorption
in the blue part), again indicating an early L spectral type.
For the gravities, $\log~g = 2$ fits best
($\log~g=4$ is better than $\log~g=0$). 
This gravity is fully consistent with the gravity-sensitive CO index
measured in our spectrum to be $0.862\pm0.035$,
yielding $\log~g = 2.5 \pm 0.8$ according to Gorlova et al. (2003).
We conclude $\log~g \simeq 2.0$ to 3.3.
The good fit indicates that the new spherically symmetric GAIA-dusty 
model is better for low gravities than the former AMES-dusty.

It is important to note that the GAIA-dusty models are applicable:
They are independant of age and stand-alone, even without interior models 
like the Baraffe et al. (2002) models. They are used, however, as outer 
boundary conditions by, e.g., Baraffe et al. (2002) to provide
their evolution models with a more realistic description
of how an interior model loses energy through the atmosphere. 
Models of the solar atmosphere usually used are also
stand-alone and independant of the interior and age.

\section{Discussion: Mass estimate for the GQ Lup companion}

At $140 \pm 50$ pc distance, with T$_{\rm eff} = 2050 \pm 450$ K and the
flux of the companion (K = $13.10 \pm 0.15$ mag), 
we can estimate its radius to be $1.2 \pm 0.5$~R$_{\rm jup}$.
With this radius and $\log~g \simeq 2.0$ to 3.3, we can estimate its mass
to be $\le 1$~M$_{\rm jup}$ 
(for $\log~g \simeq 4$ and 2~R$_{\rm jup}$, its $\sim 6$~M$_{\rm jup}$).

The Wuchterl \& Tscharnuter (2003) calculations (Fig. 4 in 
Neuh\"auser et al. 2005) indicate $\sim 1$ to 3~M$_{\rm jup}$,
co-eval with the star at $\sim 1$ Myr.

Mohanty et al. (2004a) measured gravities for isolated young
brown dwarfs and free-floating planetary mass objects. Their coolest 
objects have spectral type M7.5 and gravities as low as $\log g = 3.125$
(GG Tau Bb). This lead Mohanty et al. (2004b) to mass estimations
as low as $\sim 10$~M$_{\rm jup}$. 
GQ Lup A is younger than the Mohanty et al. Upper Sco objects.
Its companion is at least as late in spectral type, probably even cooler.
An object younger and cooler must be lower in mass.
The GQ Lup companion is fainter than the faintest Mohanty et al.
object (USco 128, $\sim 9$~M$_{\rm jup}$), so that the mass estimate 
for the GQ Lup companion is $\le 8$~M$_{\rm jup}$.

According to Burrows et al. (1997) and Baraffe et al. (2002),
the mass of the companion could be anywhere between 3 and 42~M$_{\rm jup}$,
as given in Neuh\"auser et al. (2005). However, these models are
not valid at the young age of GQ Lup, so that they are not applicable.
According to both Mohanty et al. (2004b) and Close et al. (2005),
the Baraffe et al. (2002) models overestimate the masses of young
sub-stellar objects below $\sim 30$~M$_{\rm jup}$, but underestimate
them above $\sim 40$~M$_{\rm jup}$. This is consistent with our results.

Hence, according to all valid estimations, the mass of the GQ Lup companion 
is $\sim 1$ to 8~M$_{\rm jup}$, i.e. significantly below $\sim 13$~M$_{\rm jup}$, 
hence almost certainly a planet imaged directly, to be called GQ Lup b.

{\bf Acknowledgements.} 
We would like to thank Subu Mohanty and Gibor Basri for
very fruitfull discussion about GQ Lupi b.
Especially, we would like to acknowledge Gibor Basri for
pointing us to the Mohanty et al. work.



\printindex
\end{document}